# Recurrence plots from altimetry data of some lakes in Africa

Amelia Carolina Sparavigna

Department of Applied Science and Technology, Politecnico di Torino, Torino, Italy

**Abstract** The paper shows recurrence plots obtained from time series of the level variations of four lakes in Africa (Nasser, Tana, Chad and Kainji). The data, coming from remote sensing, are provided by the United States Department of Agriculture. The recurrence plots allow a good visual comparison of the behaviours of local drainage basins.

**1. Introduction**
When represented by time series, the processes of dynamical systems can display strictly periodic on time behaviours or recurrent irregular and chaotic patterns. Intermittent chaotic behaviours can appear too. The time series, commonly used in signal processing [1], are sequences of data measured typically at successive points, which are spaced at uniform time intervals. These series can be analysed using frequency-domain and time-domain methods: the former includes spectral analysis, the latter auto-correlation and cross-correlation analysis. Among the methods based on the time-domain, we find the recurrence plots. These plots are displaying the recurrence of states of the dynamical system, that is, states that are arbitrarily close again after some time of divergence.
In 1987, Eckmann, Kamphorst and Ruelle proposed the recurrence plots to visualise the recurrence of states in phase spaces with an arbitrary number of dimensions [2,3]. These plots enabled to see the phase space trajectories through a two-dimensional representation of its recurrences. A recurrence at a different time j of a state observed at time i was depicted within a two-dimensional squared black and white image, with black dots marking recurrences. Both axes of the image were time axes. The resulting image was depending on a threshold distance, which was determining the nearer states in the phase space. Nowadays, several tools are freely available to have the recurrence plots from data files, using colours and different layouts to represent distances between the states (a list of tools is given in [4]).
The recurrence plots are used to investigate natural processes (see for instance, [5]), and for the analysis of complexity and nonlinearity of physiological, cardiac and electroencephalographic signals [6-10]. As discussed in [5], a recurrence plot is not depending on the scale factor, which we use to describe the phase space states. Then if we have to plot the recurrence of a signal, for instance a radio signal, the resulting plot will be the same for any scale factor or shift of the signal. For this reason, we can use recurrence plots to compare signals coming from different instruments, provided their origin is in the same dynamical system. One example can be the comparison of the El Niño Southern Oscillation (ENSO) and the North Atlantic Oscillation (NAO), which have their origin in the climatic earth behaviour. The comparison is important to determine climatic changes on a global scale.
In [5], the recurrence plots had been used to compare the behaviour of sunspot, solar radio data and irradiance. Here, we apply the recurrence plots to study and compare the oscillations of the levels of some lakes in Africa. The aim is to show that the recurrence plots can be useful to display differences and similarities of the local climatic behaviour of drainage basins.

## 2. Data of lake height variations

The data we will use in this paper for the recurrence plots are those provided by the United States Department of Agriculture [11]. As told in [11], the U.S. Department of Agriculture's Foreign Agricultural Service (USDA-FAS), the National Aeronautics and Space Administration and the University of Maryland are monitoring lake height variations for several large lakes around the world. The program currently utilizes radar altimeter data over inland water bodies. This monitoring is important to locate quickly the problem of regional droughts, as well as to improve the crop production estimates for those regions that are downstream from the monitored lakes. For this reason, the targeted lakes and reservoirs are located within major agricultural regions. The monitoring utilizes near-real-time radar altimeter data from an instrument on-board the Jason-2 satellite, launched in June 2008. In addition, data from the Jason-1 mission (2002-2008) and the historical data archive from TOPEX/POSEIDON mission (1992-2002) are also used [11].

The data recorded by satellites are processed after methods developed by the NASA Ocean Altimeter Pathfinder Project [11]. As told in [11], the time series of height variations are considered accurate to better than 10cm rms for the largest lakes, such as The Great Lakes, the Lake Victoria and Lake Tanganyika in Africa. Smaller lakes can expect to have accuracy better than 20cm rms. Therefore, despite some limitations, satellite radar altimeters can monitor the variation of surface height for many inland water bodies.

The web site [11] provides several time-series of water level variations by means of an interactive map of the world. The products displayed in the site are graphs and associated information in tabular form. As told in [11], for the graphs, changes in water level are real but the y-scale is arbitrary (relative) and given in meters. The x-axis refers to time with intervals of several months. Data are ranging from 1992 to 2014. In the given graphs, the top panel displays the raw height variations, while in the lower panel the same data have been smoothed.

For the following analysis by recurrence plots, we will use the data from the tabular forms. In them, some data have the overflow value 999.990. For this reason, we have determined the minimum h and the maximum H values of the level variation we can find in the table: data with value 999.990 are substituted by the average (h+H)/2. Since we aim comparing the behaviour of lakes using recurrence plots, and this analysis does not depend on scale factors, we decide to use an arbitrary scale so that the difference between h and H is the same for all the lakes. For instance, let us consider the level variations of Lake Nasser and Lake Tana. In the Figure 1, we see the smooth data from [11]. This image contains two snapshots from the web site at [11]. In the Figure 2, we see the data in arbitrary units, elaborated from the given tables as previously explained, to compare the level variations of the lakes with the recurrence plots. The same is given in the Figures 3 and 4 for Lake Chad and Lake Kainji.

## 3. Recurrence plots

The procedure to obtain the recurrence plots from the data shown in the Figures 2 and 4 is described in Reference 9. The recurrence plot $RP$ for time-discrete variables is given by:

$$RP_{i,j} = \Theta\left(\varepsilon - \|x_i - x_j\|\right) \qquad (1)$$

Here $i, j$ are the discrete time variables of the data. $x_i$ is the variable, in this case the level variation. The $\Theta$-function depends on the difference between the distance of two points and the predefined cut-off distance $\varepsilon$. We can have two-dimensional squared matrix with black and white dots as in Reference 5, or coloured images using for instance the Visual Recurrence Analysis (VRA). This is a software package for analysis, qualitative and quantitative assessment, and for nonparametric prediction of nonlinear and chaotic time series [12,13]. In the Figures 5 and 6, we display the recurrence plots in black and white as we did in [5]. Figure 5 is concerning the Lake Nasser and the Lake Tana; in the Figure 6, we have Lake Chad and Lake Kainji. The four plots had been obtained with the same threshold $\varepsilon$. We can see that the recurrence plots are characterised by a diagonal line, called the line of identity, simply meaning that each point is recurrent with itself.
If we consider two sets of data coming from the same dynamical system, we can try to display a cross recurrence plot too, defined as:

$$CRP_{i,j} = \Theta\left(\varepsilon - \|x_i - y_j\|\right) \qquad (2)$$

In a cross recurrence plots, a deviation or distortion of the diagonal part of the plot from the line of identity means that the time behaviour is different [5]. However, here we prefer to follow another approach: we prefer to compare the recurrence plots by superposing them, to have the possibility to appreciate in a colour image the similarities of the behaviours.

**4. The African Lakes**
From the Figures 5 and 6 we can appreciate that we are not observing a strictly periodic signal, such as a sine or cosine wave. In fact, if we compare these plots with those shown in [14], we see that the lake Kainji recurrence plot (Figure 6) is similar to the recurrence plot of the chaotic signal of a Rossler attractor. This similarity can be greatly appreciated, if we compare the image concerning this attractor given in [14], with the images in the Figure 7 where we used the Visual Recurrence Analysis (VRA) for the data of the lake. As told in [14], the Rossler attractor displays a global structure with local instabilities. It has some similarities to the Lorenz attractor [15], obtained in 1963 by Edward Lorenz, which is an attractor displayed by a simplified mathematical model for atmospheric convection [16].
Let us compare the behaviour of the lakes. In the Figure 2 and 5, data from Lake Nasser and Lake Tana are shown together. In fact, these lakes have in common the same water. Lake Nasser is created by the River Nile, and the 90% of water of this river is sourced by the Blue Nile, which springs from the Lake Tana [17]. The flow of the Blue Nile varies considerably over its yearly cycle and is the main contribution to the large natural variation of the Nile flow. The flow of the Blue Nile is controlled by the Roseires Dam; downstream, at Khartoum, the Blue Nile and White Nile rendezvous becoming one great river. The Nile's path then continues in Sudan. The river is again controlled by the Merowe Dam and his basin [18,19]. Then the Nile arrives to the Aswan Dam, where we find the huge Nasser Lake. The Lake Nasser and the Lake Tana are then connected: in the Figure 8 we compare the recurrence plots of data from 2009-2012. The panel on the right is showing the superposition of the images: in red, we have the data from Lake Tana and in black, those from Lake Nasser. Both have a "butterfly" texture, time shifted. If we look at Figure 5, we can see a similar behaviour of data from 1992-2002.
In the Figure 9, we are doing the same for the Lake Chad and the lake Kainji. These two lakes are important freshwater and fish resources for the local populations [20]. We have already observed that the recurrence plot of the lake Kainji is similar to that of the Rossler attractor: in the Figure 9, we see some similarities in the behavior of the level of the two lakes, ruled by the local rain season.

**5. Conclusion**
We have proposed in this paper the recurrence plots of four time-series data concerning level variations of four lakes in Africa. From the plots, we can see the oscillations of the level in them. One of the lakes, the Lake Kainji has a recurrence plot similar to that of a Rossler attractor. As already observed in Reference 5, the recurrence plots are like "fingerprints" of the dynamical behaviour of the system, able to give an image of recurrences of maxima and minima. As we can see from the images given in this paper, the information provided by these "fingerprints" turns out to be quite clear, and suitable for a quick comparison of different sets of data.

**References**
1. G. Box and G. Jenkins, Time Series Analysis Forecasting and Control, Holden-Day, San Francisco, 1976.
2. J.-P. Eckmann, S.O. Kamphorst and D. Ruelle, Recurrence Plots of Dynamical Systems, Europhysics Letters, 1987, Volume 5, Pages 973-977.
3. J.-P. Eckmann and D. Ruelle, Ergodic Theory of Chaos and Strange Attractors, Review of Modern Physics, 1985, Volume 57, Issue 3, Pages 617-656.
4. http://www.recurrence-plot.tk/
5. A.C. Sparavigna, Recurrence Plots of Sunspots, Solar Flux and Irradiance, arXiv:0804.1941 [physics.pop-ph], 2008.
6. C.L. Webber Jr and J.P. Zbilut, Dynamical Assessment of Physiological Systems and States Using Recurrence Plot Strategies, Journal of Applied Physiology, 1994, Volume 76, Issue 2, Pages 965-973.
7. J.P. Zbilut, N. Thomasson, and C.L. Webber, Recurrence Quantification Analysis as a Tool for Nonlinear Exploration of Nonstationary Cardiac Signals, Medical Engineering & Physics, 2002, Volume 24, Issue 1, Pages 53-60.
8. Gaoxiang Ouyang, Li Xiaoli, Dang Chuangyin and D.A. Richards, Using Recurrence Plot for Determinism Analysis of EEG Recordings in Genetic Absence Epilepsy Rats, Clinical Neurophysiology, 2008, Volume 119, Issue 8, Pages 1747-1755.
9. N. Marwan and J. Kurths, Cross Recurrence Plots and Their Applications, in Mathematical Physics Research at the Cutting Edge, C.V. Benton Editor, pp.101-139, Nova Science Publishers, 2004.
10. J.P. Zbilut and C.L. Webber Jr., Embeddings and Delays as Derived from Quantification of Recurrence Plots, Phys. Letters A, 1992, Volume 171, Issue 3-4, Pages 199-203.
11. USDA, at www.pecad.fas.usda.gov/ cropexplorer/ global_reservoir/
12. E. Kononov, Visual Recurrence Analysis, www.visualization-2002.org/
13. E. Kononov, visual-recurrence- analysis.software.informer.com/4.9/
14. E. Kononov, www.visualization-2002.org/ VRA_Main_Description_.html
15. Vv. Aa., Wikipedia, /wiki/Rössler_attractor
16. Vv. Aa., Wikipedia, /wiki/Lorenz_system
17. Vv. Aa., Wikipedia, /wiki/Nile#Blue_Nile
18. A.C. Sparavigna, The Merowe Dam on the Nile, Archaeogate, 3 December 2012, available at http://porto.polito.it/2379391/
19. A.C. Sparavigna, Merowe Dam and the Inundation of Paleochannels of the Nile, arXiv:1011.4911 [physics.geo-ph], 2010.
20. G.N. Shimang, Post-Harvest Losses in Inland Fisheries in Nigeria with Emphasis on Lake Chad and Lake Kainji, Proceedings of the Symposium on Post-Harvest Fish Technology, Cairo, Egypt, 21–22 October, 1990.

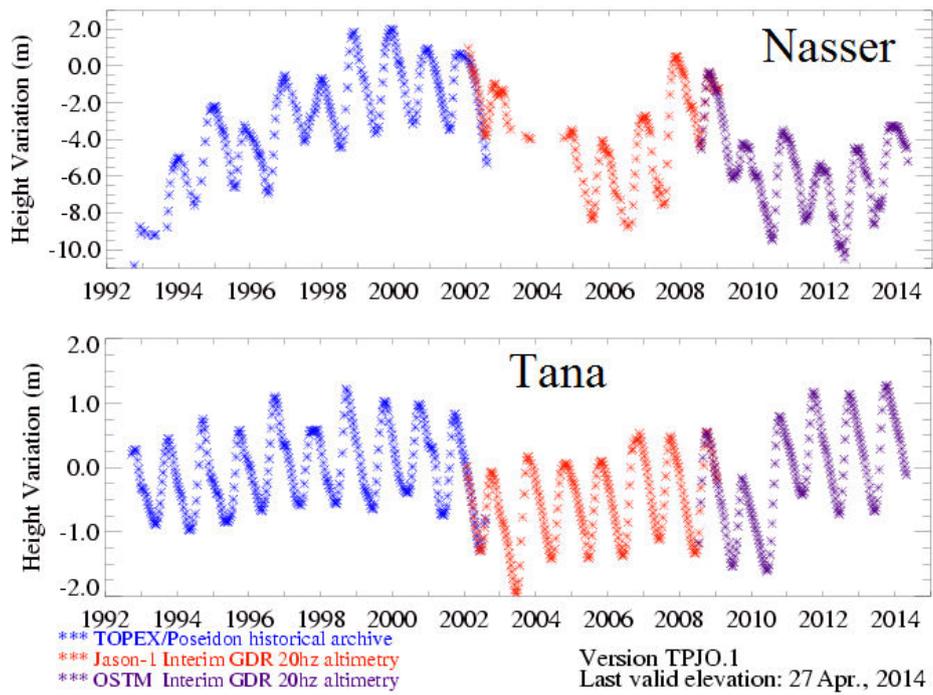

figure 1 – The height variations of the Lake Nasser and of the Lake Tana. The images are snapshots of graphs representing the smoothed data from satellites, reported in [11] by USDA.

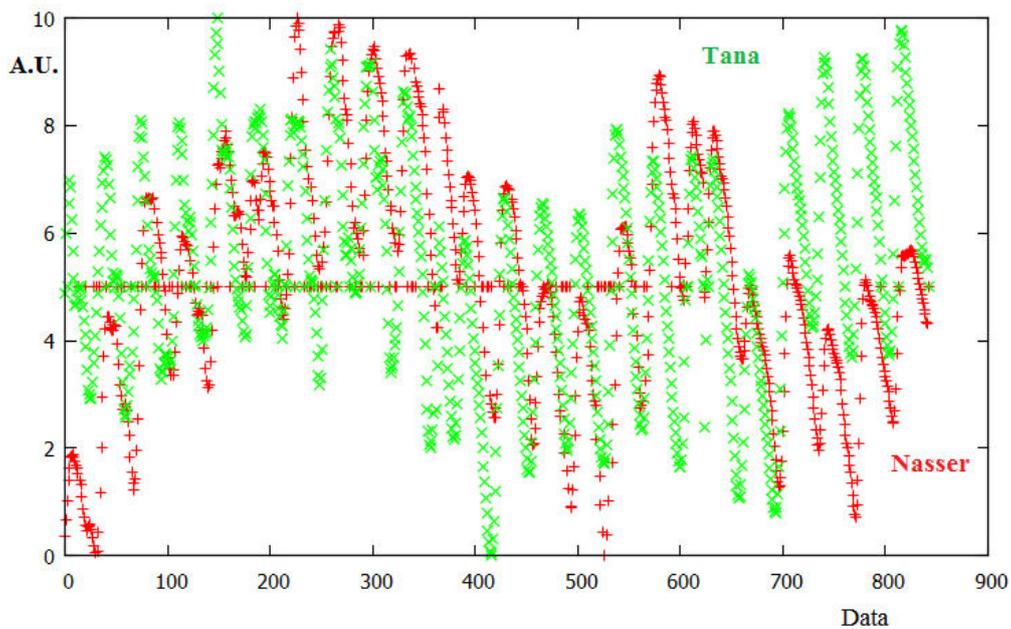

Figure 2 – The height variations of Lake Nasser (red) and Lake Tana (green). The units on y-axis are arbitrary. The data used are those in the Figure 1, processed as explained in the text.

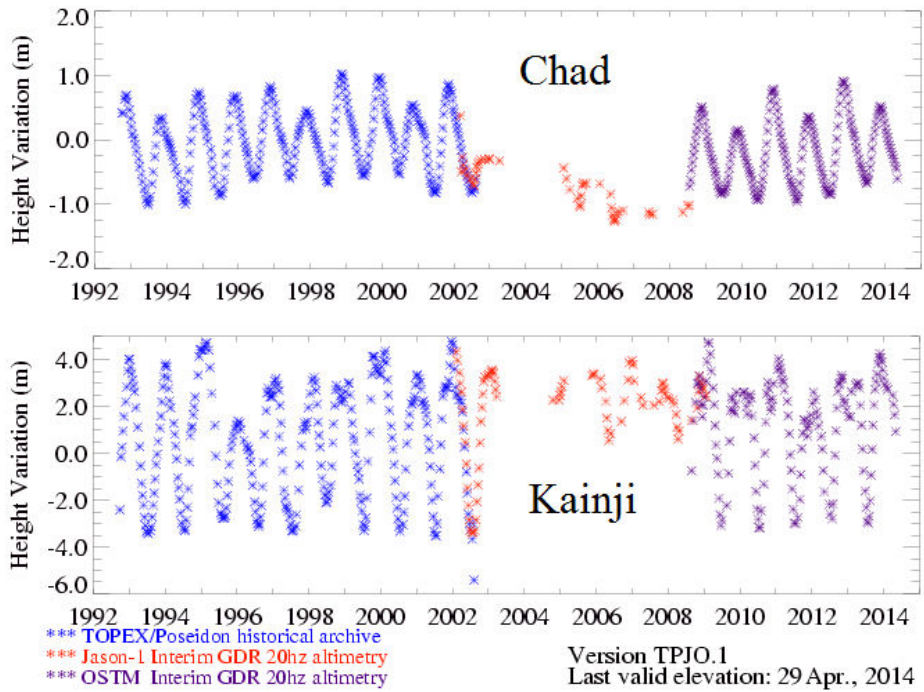

Figure 3 – The height variations of the Lake Chad and of the Lake Kainji. The images are snapshots of graphs representing the smoothed data from satellites, reported in [11] by USDA.

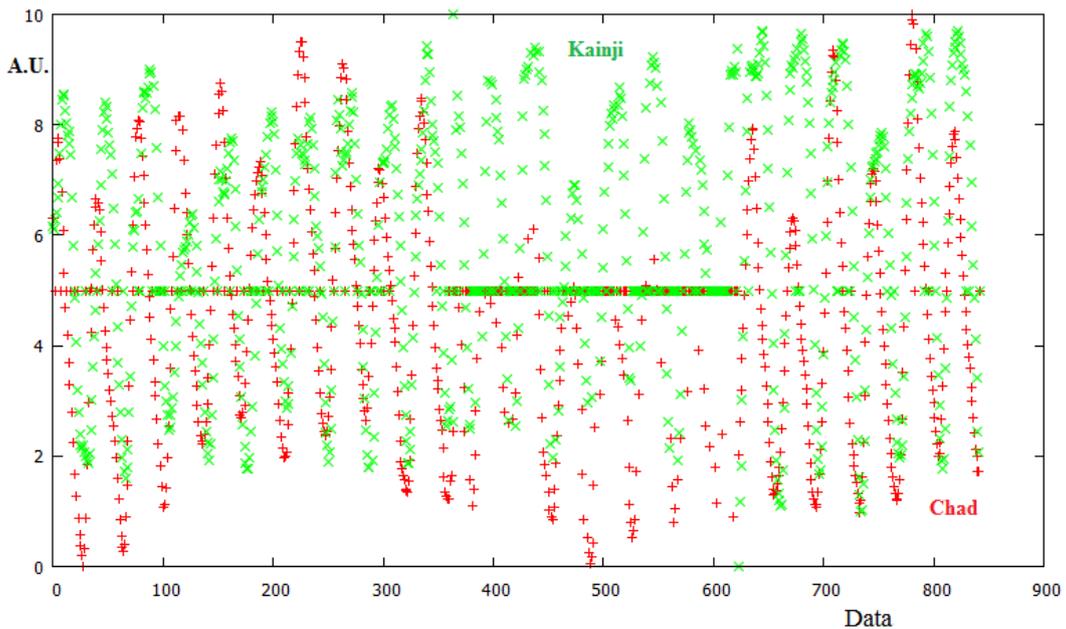

Figure 4 – The height variations of Lake Chad (red) and Lake Kainji (green). The units on y-axis are arbitrary. The data used are those in the Figure 3, processed as explained in the text.

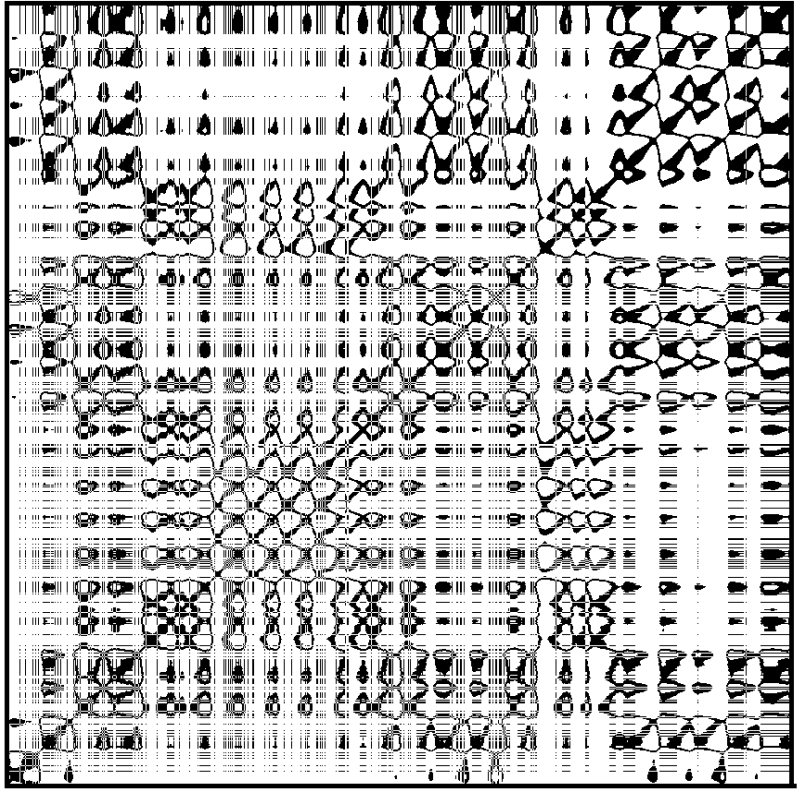
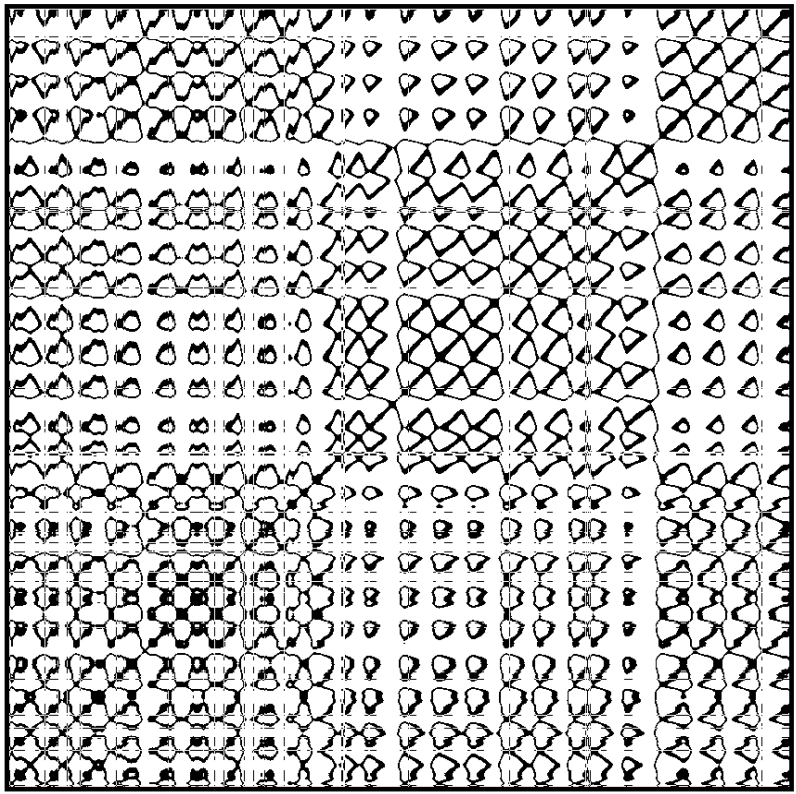

Figure 5 - Recurrence plots of the time series in the Figure 2 of Lake Nasser and Lake Tana.

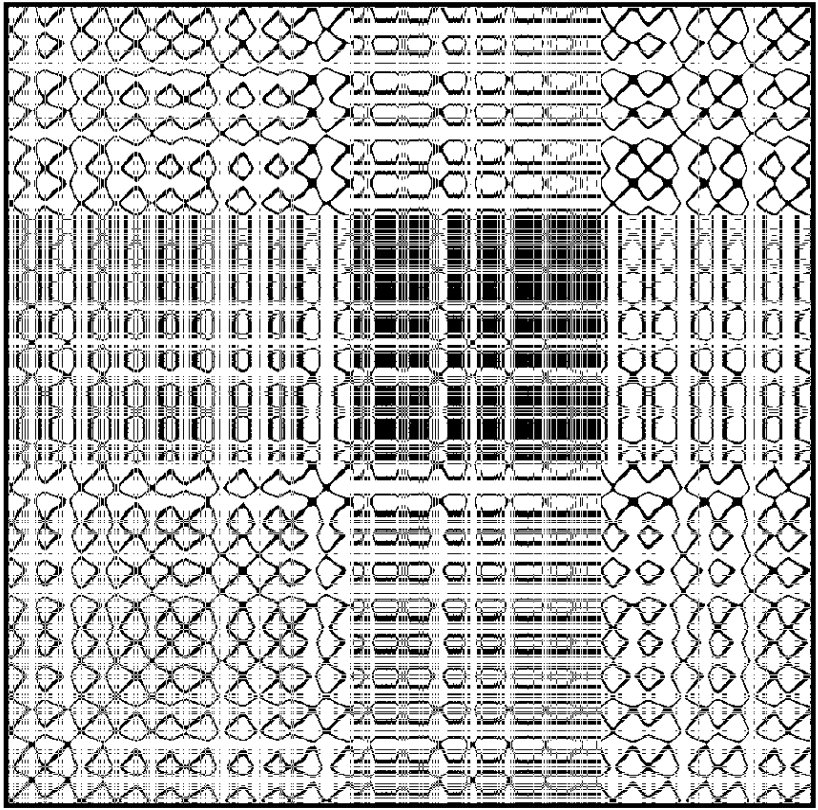

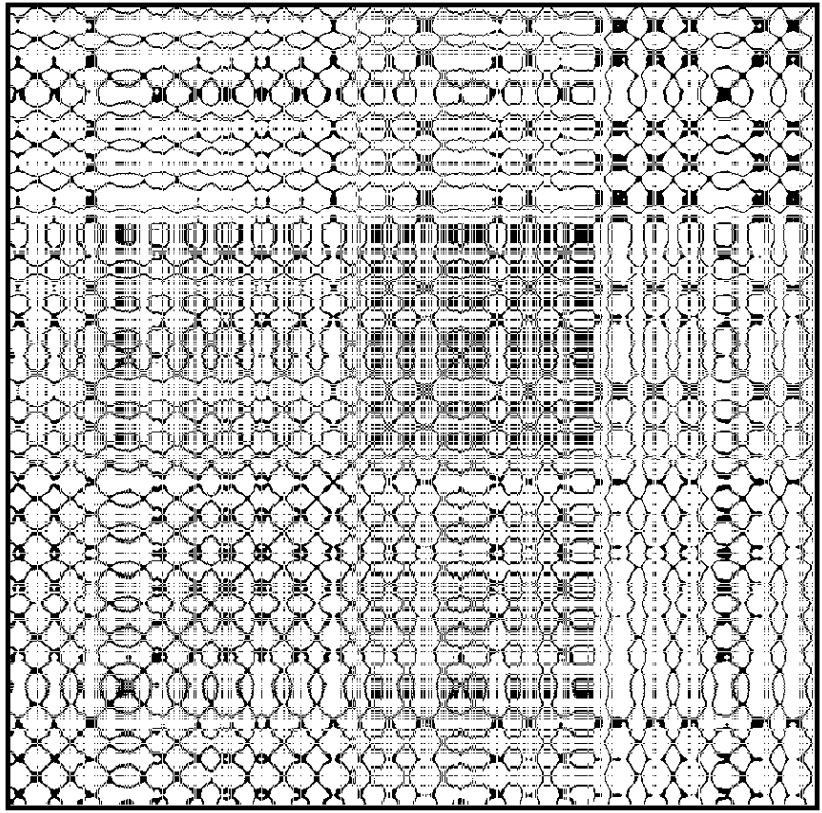

Figure 6 - Recurrence plots of the time series in the Figure 4 of the Lake Chad and Lake Kainji.

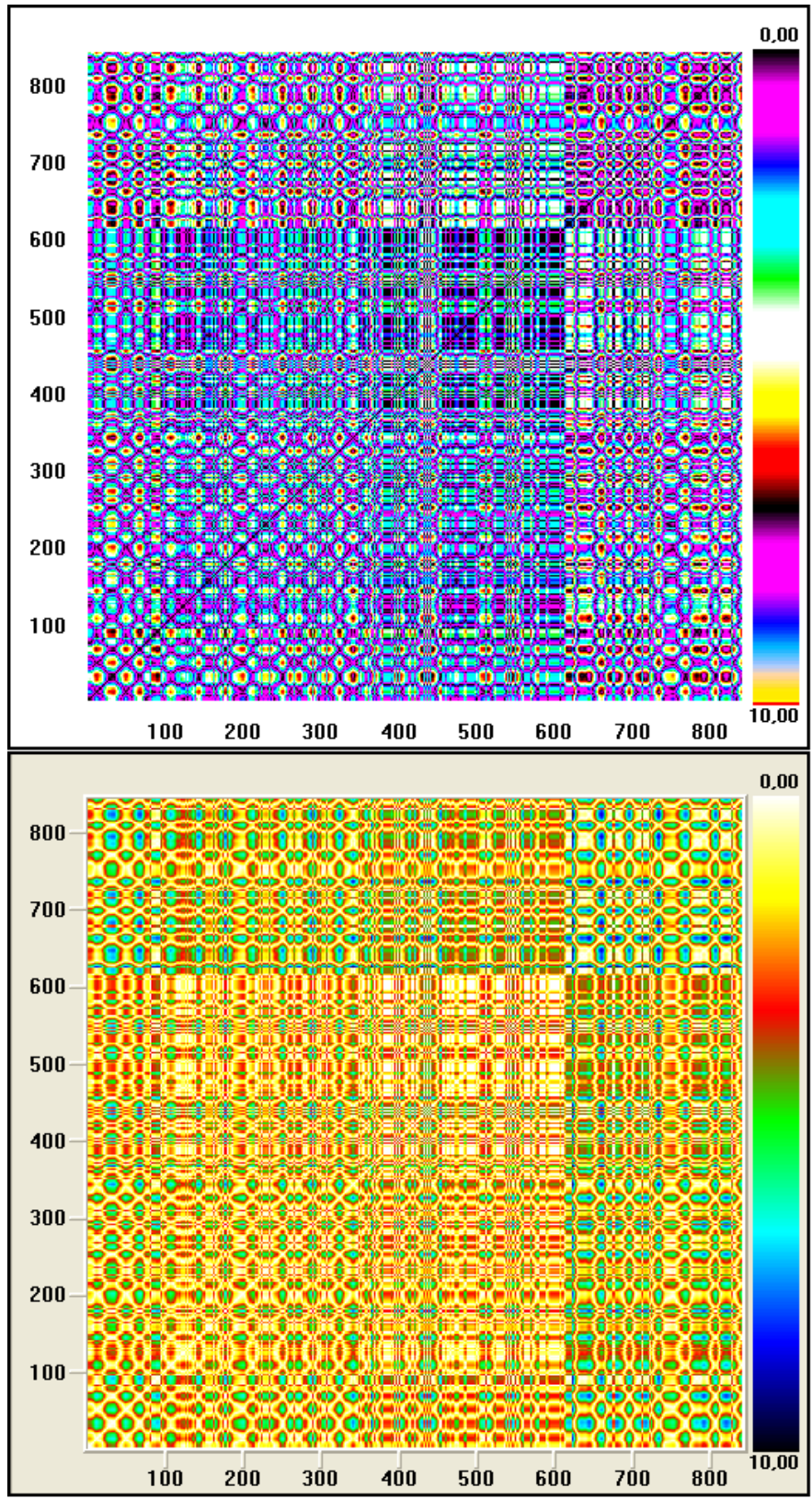

Figure 7 - Recurrence plots of the time series of Lake Kainji, obtained by means of the Visual Recurrence Analysis (VRA), in two different layouts [12-14]. If we compare these plots with those shown in [14], we see that the lake Kainji recurrence plot is similar to the recurrence plot of the chaotic signal of a Rossler attractor. This attractor displays a global structure with local instabilities. It is similat to the Lorenz attractor [15], coming from a mathematical model for atmospheric convection [16].

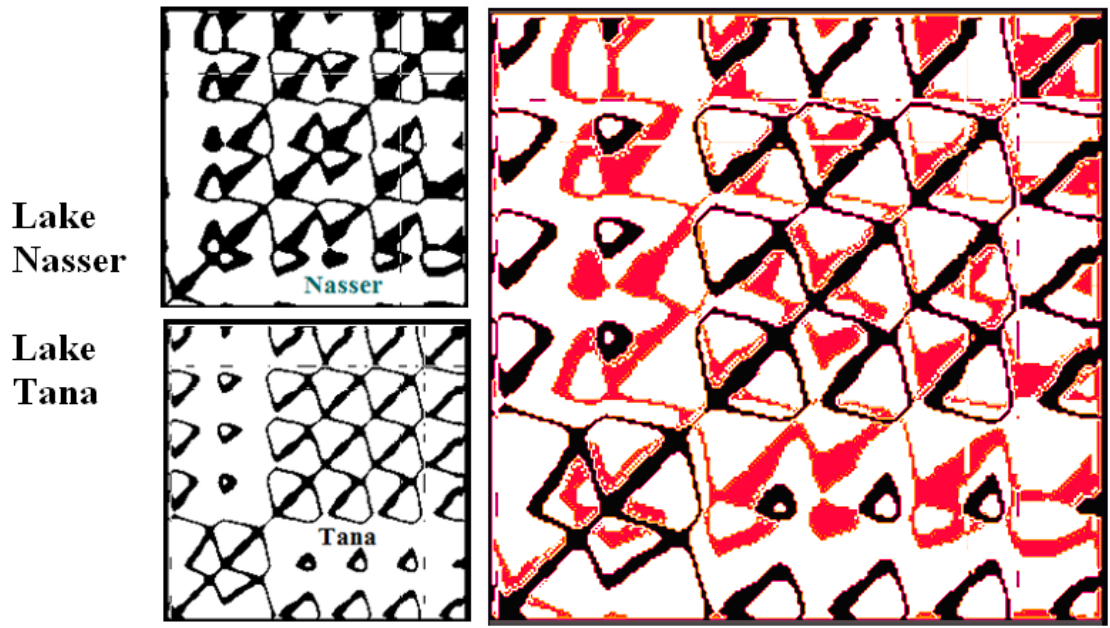

Figure 8 – Comparison of the recurrence plots of data from 2009-2012 of Lake Nasser and Lake Tana. In the right panel, we have a superposition the images: we see the data from Lake Tana in red, and in black those from Lake Nasser. We see the same "butterfly" behaviour with a time shift.

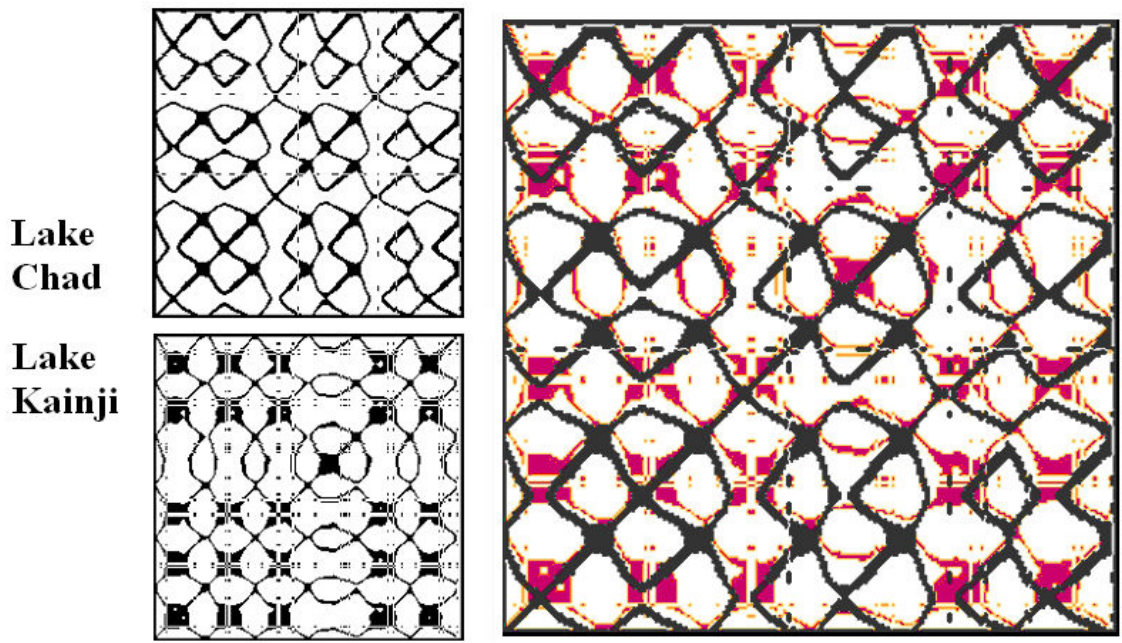

Figure 9 – Comparison of the recurrence plots of data from 2009-2012 of Lake Chad and Lake Kainji. In the right panel, we have the superposition of images: the data from Lake Kainji are in red, and in black those from Lake Chad. Again, we have a similar behaviour with a time shift.